\documentclass[prb,twocolumn,showpacs,preprintnumbers,amsmath,amssymb,citeautoscript]{revtex4-1}

\usepackage{graphicx}
\usepackage{dcolumn}
\usepackage{bm}
\usepackage{color}

\newlength{\figwidth}
\def\sldmit{EtMe$_3$Sb[Pd(dmit)$_2$]$_2$}

\begin{document}

\title{Novel Pauli-paramagnetic quantum phase in a Mott insulator}

\author{D.~Watanabe$^{1,*}$}
\author{M.~Yamashita$^{1,2}$}
\altaffiliation{These authors contributed equally to this work. Correspondence and requests for materials should be addressed to T.S. (email: shibauchi@scphys.kyoto-u.ac.jp) or to
Y.M. (email: matsuda@scphys.kyoto-u.ac.jp).}
\author{S.~Tonegawa$^1$}
\author{Y.~Oshima$^2$}
\author{H.\,M.~Yamamoto$^{2,3}$}
\author{R.~Kato$^2$}
\author{I.~Sheikin$^4$}
\author{K.~Behnia$^5$}
\author{T.~Terashima$^6$}
\author{S.~Uji$^6$}
\author{T.~Shibauchi$^1$}
\author{Y.~Matsuda$^1$}

\affiliation{$^1$Department of Physics, Kyoto University, Kyoto 606-8502, Japan}
\affiliation{$^2$RIKEN, Wako, Saitama 351-0198, Japan}
\affiliation{$^3$Institute for Molecular Science, Myodaiji, Okazaki 444-8585, Japan}
\affiliation{$^4$Grenoble High Magnetic Field Laboratory, CNRS, 38042 Grenoble Cedex 9, France}
\affiliation{
$^5$LPEM (CNRS-UPMC), Ecole Sup\'{e}rieure de Physique et de Chimie Industrielles, 75005 Paris, France}
\affiliation{$^6$National Institute for Materials Science, Ibaraki 305-0003, Japan}

\date{\today}

\begin{abstract}
{\bf
In Mott insulators, the strong electron-electron Coulomb repulsion prevents metallicity and charge excitations are gapped. In dimensions greater than one, their spins are usually ordered antiferromagnetically at low temperatures.  Geometrical frustrations can destroy this long-range order, leading to exotic quantum spin liquid (QSL) states~\cite{Balents10}.  However, their magnetic ground states have been a long-standing mystery~\cite{Anderson73,Moessner01,Wen02,Morita02,Motrunich05,Lee05,Yoshioka09,Qi09,Block11,Pot12,Mis99,Kyu06,Lawler08}. Here we show that a QSL state in the organic Mott insulator EtMe$_3$Sb[Pd(dmit)$_2$]$_2$ with two-dimensional triangular lattice~\cite{Itou08,Yamashita10,Itou10,Yamashita11,Kanoda11} has Pauli-paramagnetic-like low-energy excitations, which are a hallmark of itinerant fermions.  Our torque magnetometry down to low temperatures (30\,mK) up to high fields (32\,T) reveal distinct residual paramagnetic susceptibility comparable to that in a half-filled two-dimensional metal. This demonstrates that the system is in a magnetically gapless ground state,  a critical state with infinite magnetic correlation length.  Moreover, our results are robust against deuteration,  pointing toward the emergence of an extended `quantum critical phase', in which low-energy spin excitations behave as in paramagnetic metals with Fermi surface, despite the frozen charge degree of freedom.
}
\end{abstract}

\maketitle

At sufficiently low temperatures, condensed matter systems generally tend to order.    QSLs are a prominent exception, in which no local order parameter is formed while the entropy vanishes at zero temperature.   In two- or three-dimensional (2D or 3D) systems, it is widely believed \cite{Balents10} that QSL ground states may emerge when interactions among the magnetic degrees of freedom are incompatible with the underlying crystal geometry.  Typical 2D examples of such geometrically frustrated systems can be found in triangular and kagome lattices.  Largely triggered by the proposal of the resonating-valence-bond theory on spin-1/2 degrees of freedom residing on a 2D triangular lattice \cite{Anderson73} and its possible application to high-transition-temperature superconductivity \cite{Lee06},  realizing/detecting QSLs has been a long-sought goal.   Until now, quite a number of QSLs with various types of ground states have been proposed \cite{Anderson73,Moessner01,Wen02,Morita02,Motrunich05,Lee05,Yoshioka09,Qi09,Block11,Pot12,Mis99,Kyu06,Lawler08}, but the lack of real materials had prevented us from understanding the nature of QSLs.

The two recently discovered organic Mott insulators,  $\kappa$-(BEDT-TTF)$_2$Cu$_2$(CN)$_3$ \cite{Shimizu06,Yamashita08,Yamashita09,Goto10,Pratt11} and EtMe$_3$Sb[Pd(dmit)$_2$]$_2$ \cite{Itou08,Yamashita10,Itou10,Yamashita11} (see Fig.\:1a), are very likely to be the first promising candidates to host a QSL in real bulk materials \cite{Kanoda11}.   In these systems, (BEDT-TTF)$^{\frac{1}{2}+}$ cations or  Pd(dmit)$_2^{\frac{1}{2}-}$ anions are strongly dimerized and spin-1/2 units of (BEDT-TTF)$_2$ and [Pd(dmit)$_2$]$_2$ form 2D triangular structure. In both systems, nuclear magnetic resonance (NMR)\cite{Shimizu06,Itou08} and muon spin rotation\cite{Goto10,Pratt11} measurements exhibit no sign of long-range magnetic ordering down to a very low temperature whose energy scale corresponds to $J/10,000$ ($J/k_B\sim 250$\,K for both compounds), where $J$ is the nearest-neighbour exchange coupling energy.  These findings aroused great interest because even in a quantum spin-1/2 triangular lattice antiferromagnet, the frustration brought on by the nearest-neighbour Heisenberg interaction is known to be insufficient to destroy the long-range ordered ground state.   Several classes of QSL states have been put forth to explain these exotic spin states, yet their ground state remains puzzling.   Between the two compounds, EtMe$_3$Sb[Pd(dmit)$_2$]$_2$ appears to be  more ideal to single out genuine features of the QSL,  because  more  homogeneous QSL state can be attained at low temperatures \cite{Itou10}.  In contrast,  $\kappa$-(BEDT-TTF)$_2$Cu$_2$(CN)$_3$ has been reported to have an inhomogeneous and phase-separated spin state \cite{Shimizu06,Goto10}.


\begin{figure*}[ht]
\includegraphics[width = 0.9\linewidth]{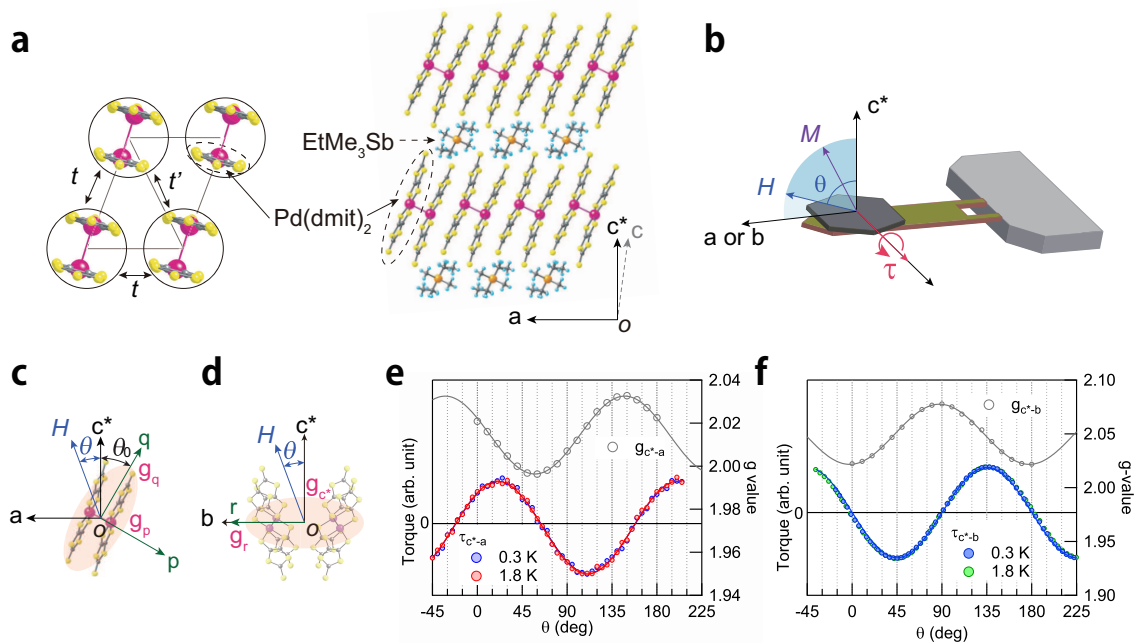}
\caption{{\bf Magnetic torque measurements in the QSL state of the organic Mott insulator with 2D triangular lattices.} 
{\bf a}, The crystal structure of {\sldmit} that viewed from the $b$ axis (right).  The crystal consists of 2D layers of Pd(dmit)$_2$ molecules separated by the non-magnetic cation EtMe$_3$Sb.   The $c^\ast$ axis is defined as a perpendicular axis to the $ab$ plane.   In the Pd(dmit)$_2$ layer, the face-to-face pair of Pd(dmit)$_2$$^\frac{1}{2}$ anions are strongly dimerised (encircled) and  spin-1/2 (one electron) units of encircled [Pd(dmit)$_2$]$_2$ form the triangular lattice.   The ratio of the nearest-neighbour inter-dimer transfer integrals is estimated to be $t'/t=0.92$ from the tight-binding calculations based on the crystallographic data at room temperatures, indicating a nearly-ideal triangular structure~\cite{Kanoda11}. 
{\bf b}, Schematic drawing of torque measurements using a cantilever. By rotating magnetic field from the $c^\ast$ axis to $a$ or $b$ axis, the magnetic torque component perpendicular to the rotating field plane is measured as a function of the polar angle $\theta$. 
{\bf c}, {\bf d},  The gyromagnetic ratio ($g$-factor) of  [Pd(dmit)$_2$]$_2$ with spin-1/2.  The principal axes of $g$-tensor denoted by $p$, $q$ and $r (\parallel b)$ do not coincide with crystallographic axes.   The spheroidal pink shaded regions represent the anisotropy of $g$-factor (Supplementary Information). 
{\bf e,f },  Angular variation of magnetic  torque  (filled circles, left axis) and the $g$-factor determined by the electron paramagnetic resonance at 4.2\,K (open circles, right axis) when the magnetic field is rotated within the $c^\ast-a$ plane ({\bf e}) and   $c^\ast-b$ plane ({\bf f}) . The solid lines are fits to the $\sin2\theta$ dependence.}
\end{figure*}


What is remarkable in the QSL state of  EtMe$_3$Sb[Pd(dmit)$_2$]$_2$ is the presence of the gapless excitations, which are highly mobile with long mean free path, as reported by thermal conductivity \cite{Yamashita10} and  heat capacity measurements \cite{Yamashita11}. However, there are two fundamental questions that still remain open. The first one is the magnetic nature of the gapless excitations, which cannot be obtained from the above measurements.  Whether the excitations are magnetic or nonmagnetic bears direct implications on the spin-spin correlation function.  The second question concerns the phase diagram; how the QSL varies when tuned by  the non-thermal parameter, such as frustration, is the key to understanding the ground state.  The uniform susceptibility in the low-temperature limit provides  pivotal information on the magnetic character of the ground state.  However,  conventional measurements using SQUID magnetometer are susceptible to paramagnetic contributions due to impurities.  Indeed, for EtMe$_3$Sb[Pd(dmit)$_2$]$_2$, the temperature region of  importance  is below $\sim$5~K, but SQUID signal is dominated by the paramagnetic contributions residing on the non-magnetic cation EtMe$_3$Sb layers \cite{Yamashita10}.   To resolve the intrinsic magnetic susceptibility of the QSL in single crystalline samples, we performed the magnetic torque $\bm{\tau}$ measurements by using micro-cantilever method \cite{Okazaki11} (Fig.\:1b), which is the cross product of the applied field $\bm{H}$ and the magnetization $\bm{M}$ ($M_i=\Sigma_i\chi_{ij}H_j$ where $\chi_{ij}$ is the spin susceptibility tensor diagonalised along the magnetic principal axes, see Figs.\:1c and d) as $\bm{\tau} = \mu_0 V \bm{M} \times \bm{H}$, where $\mu_0$ is the permeability of vacuum and $V$ is the sample volume.   The most advantage of this method is that the large Curie contribution from impurity spins is cancelled out (Supplementary Information).


\begin{figure*}[htb]
\includegraphics[width=0.8\linewidth]{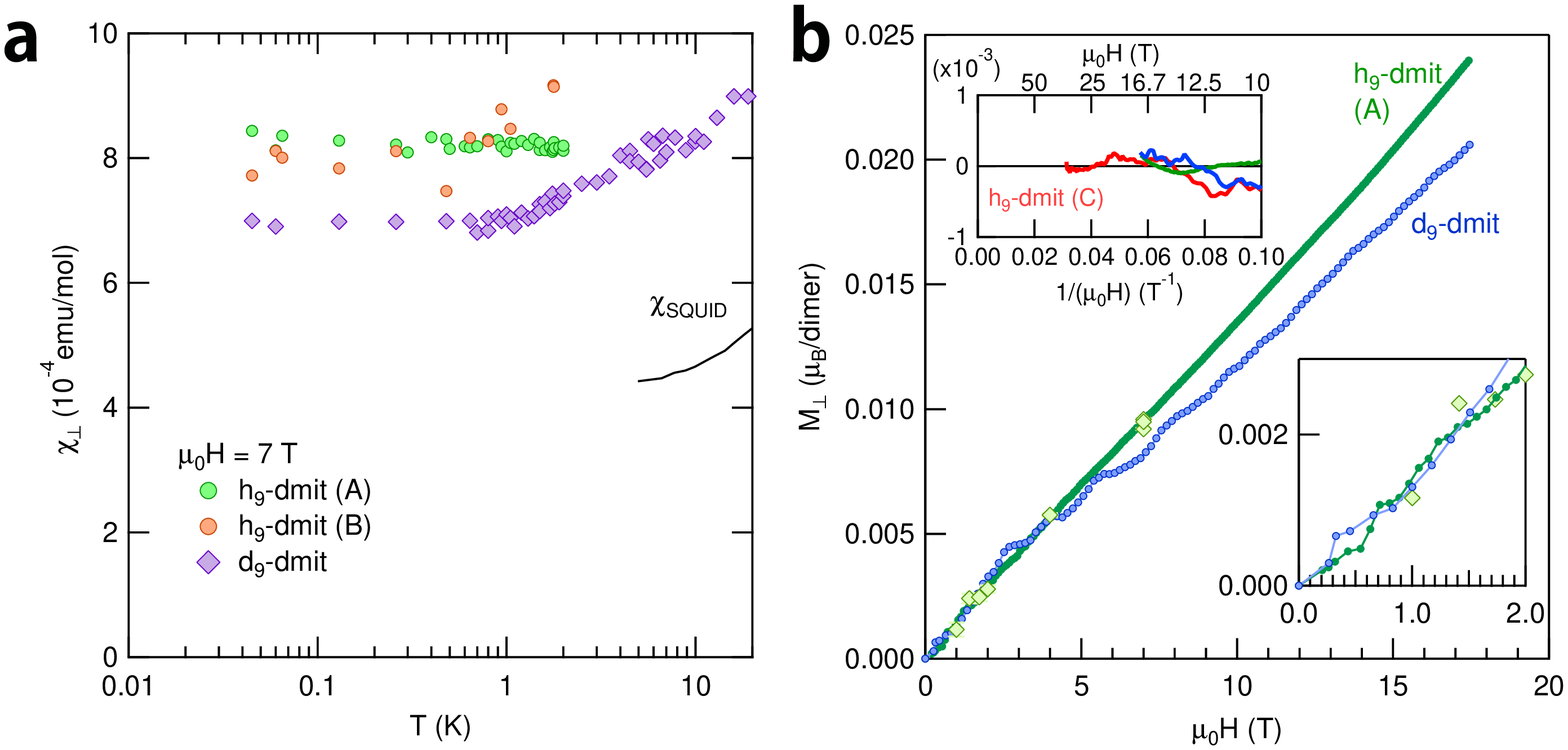} 
\caption{
{\bf Magnetic susceptibility and magnetization in the QSL state of {\sldmit} and its deuterated compound.}
{\bf a}, Temperature dependence of $\chi_\perp$ of two crystals of pristine material (h$_9$-dmit (A), h$_9$-dmit (B)) and a deuterated sample (d$_9$-dmit) determined by the torque curves at 7\,T.  The solid line is the susceptibility data of polycrystalline sample measured by the SQUID magnetometer \cite{Itou08}, where the large  Curie contribution from impurities has been subtracted.
{\bf b}, Field dependence of magnetization $M_\perp$ determined from $\chi_\perp$ of pristine (h$_9$-dmit (A), h$_9$-dmit (C)) and deuterated (d$_9$-dmit) samples. The data shown as filled circles ($T = 30$\,mK) were taken by sweeping field at a fixed angle $\theta$. The light green diamonds show the amplitude of the whole torque curves $\tau(\theta)$ at 50\,mK for h$_9$-A sample. The lower inset shows $M_\perp$ in the  low field region.   The upper  inset is the deviation from the high-field linear dependence plotted against the inverse field. The data up to 35\,T (for h$_9$-dmit (C)) show no apparent quantum oscillations.
} 
\end{figure*}


Figures\:1e and f depict the magnetic torque curves of  EtMe$_3$Sb[Pd(dmit)$_2$]$_2$ (h$_9$-dmit) when the magnetic field of $\mu_0H=7$\,T is rotated within the $ac^{\ast}$ and $bc^{\ast}$ plane, where $\theta$ is the polar angle measured from the $c^{\ast}$ axis.   In these figures, we also plot the angular dependence of  the gyromagnetic ratio ($g$-factor) arising from  the spin-1/2 of [Pd(dmit)$_2$]$_2$ dimers.  The uniform susceptibility $\chi_{\perp}$ and magnetization $M_{\perp}$ perpendicular to the $ab$ plane ($\parallel c^{\ast}$ axis) can be determined  from the torque and $g$-factors (Supplementary Information).  Figures\:2a and b depict the temperature dependence of $\chi_{\perp}$ at $\mu_0H=7$\,T and the field dependence of $M_{\perp}$ at $T=30$\,mK, respectively, for pristine crystal (h$_9$-dmit) and its deuterated compound (d$_9$-dmit), which has different degrees of geometrical frustration \cite{Yamashita11}. The absolute value of $\chi_{\perp}$ is nearly 1.5 times larger than $\chi$ of polycrystalline samples of h$_9$-dmit determined by the SQUID magnetometer (Fig.\:2a).

The most notable features are (i) $\chi_{\perp}$  is  temperature independent below 2\,K and remains finite in zero temperature limit (Fig.\:2a) and (ii) $M_{\perp}$ increases nearly linearly with the applied field up to 17\,T (Fig.\:2b).  In the dmit materials the orbital paramagnetic (Van Vleck) susceptibility is very small (Supplementary Information), and thus the spin (Pauli) paramagnetism should be responsible for the observed temperature-independent and field-linear paramagnetic response. These results in the low-temperature limit provide direct evidence of the presence of low-lying gapless `magnetic' excitations.  It is also clear that the gapless excitations observed in the thermal conductivity~\cite{Yamashita10} and the heat capacity~\cite{Yamashita11}  measurements contain the magnetic ones.   As the spin gap $\Delta$ is inversely proportional to the magnetic correlation length $\xi$, this spin-gapless QSL state has infinite $\xi$.  Therefore the presence of the gapless magnetic excitations is of crucially importance, as it immediately indicates that the QSL is in a critical state where the spin-spin correlation function decays with distance $r$ in an anomalous non-exponential manner: a simple example is the power-law decay $\langle S^z(r)S^z(0)\rangle \propto r^{-\eta}$ with $\eta>0$ for the so-called algebraic spin liquid \cite{Wen02}.  While NMR measurements report an anomaly at 1\,K \cite{Itou10}, no anomaly is observed in the magnetic susceptibility, which is consistent with the thermal conductivity \cite{Yamashita10} and specific heat measurements \cite{Yamashita11}.  This indicates that the NMR anomaly at 1\,K is not a signature of the thermodynamic transition, but a phenomenon which may be related to the slow spin dynamics of the QSL.  


\begin{figure*}[htb]
\includegraphics[width=0.8\linewidth]{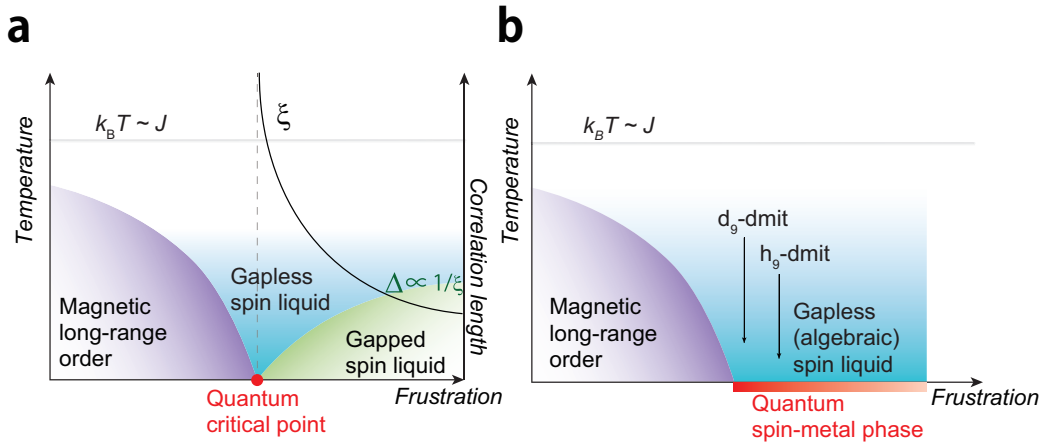} 
\caption{
{\bf Schematic phase diagrams of temperature versus frustration for Mott insulating antiferromagnets with 2D triangular lattice.} 
The QSL appears in two distinctly different ways.  
{\bf a}, The ordered ground state (with no gap) immediately evolves into a gapped QSL. A gapless QSL, in which  the spin correlation decays algebraically with distance,  appears only at the QCP, which is the border between the ordered state and the gapped QSL state.  In this case, the observation of the critical gapless QSL requires fine tuning of the material parameters. 
{\bf b}, An extended quantum critical `phase', where the gapless QSL with infinite correlation length is present, appears beyond a critical value.   The fact that both h$_9$-dmit and d$_9$-dmit with different degrees of frustration exhibit essentially the same paramagnetic behaviours with no spin gap indicates that both materials are in the same critical state.  This leads us to conclude the presence of quantum critical phase ({\bf b}),  rather than a quantum critical point ({\bf a}), in EtMe$_3$Sb[Pd(dmit)$_2$]$_2$.
}
\end{figure*}


There are two possible phase diagrams for the QSL as a function of frustration, as illustrated in Figs.\:3a and b.   With increasing frustration, a magnetically ordered phase with broken symmetry is destroyed at a critical value, beyond which a QSL phase without broken symmetry appears.   The first case is that a gapless QSL with infinite $\xi$ emerges only at a quantum critical point (QCP)~\cite{Sachdev99}, beyond which the spin-gapped QSL with finite $\xi$ appears (Fig.\:3a).  In this case the QSL is placed in a category of topological spin liquid in which spin correlation function decays exponentially with distance as $\exp(-r/\xi)$ except at QCP. 
The second case is that there is an extended critical `phase', where the gapless QSL with infinite correlation length is stably present (Fig.\:3b).  These two cases can be distinguished by comparing $\chi_{\perp}$ of h$_9$-dmit and d$_9$-dmit at $k_B T \ll J$, because deuteration of the Me groups  in EtMe$_3$Sb[Pd(dmit)$_2$]$_2$ changes the degrees of geometrical frustration by reducing $t'/t$~\cite{Yamashita11}. As shown in Figs.\:2a and b, both h$_9$- and d$_9$-dmit crystals exhibit essentially the same paramagnetic behaviours with no spin gap, indicating that both materials  are in a critical state down to $k_B T \sim J/10,000$. These results lead us to conclude the presence of an extended quantum critical phase of the spin-gapless QSL (Fig.\:3b).

The presence of a stable QSL `phase' with gapless magnetic excitations is hard to explain in the conventional bosonic picture. Within this picture, when the magnetic long-range order is established, the gapless Nambu-Goldstone bosons, i.e. magnons, would arise as the consequence of the spontaneous symmetry breaking of the spin SU(2) symmetry. When the system evolves into the QSL state with no symmetry breaking, the common wisdom is that both of the magnetic order and Nambu-Goldstone bosons disappear at the same time. This implies that the presence of gapless QSL phase has some exotic elementary excitations.  
Among them,   `spinon' excitation, which is a chargeless spin-1/2 quasiparticle (half of the magnons), has been discussed extensively. In theories of the QSL with the bosonic spinons, the magnetic order is described by the Bose-Einstein condensation (BEC)~\cite{Giamarchi2008} of the bosonic spinons. When the magnetic order is destroyed or the BEC disappears, the bosonic spinons in the resulting QSL can only be gapless at the critical point beyond which they become gapped~\cite{Qi09}. The presence of an extended region of spin-gapless QSL phase at zero or very low temperatures, therefore, would suggest that the underlying QSL phase may be better described by the QSL with fermionic spinons~\cite{Lee05,Block11}.

Indeed,  the gapless magnetic excitations bears some resemblance to the elementary excitations in the spin channel of metals with Fermi surface, i.e. Pauli paramagnetism.   Here we examine a thermodynamic test simply by analysing the data in accordance with the assumptions that the elementary quasiparticles in the QSL phase are 2D fermions with Fermi surface, even though the system is an insulator (see Supplementary Information for detail calculations). Using $\chi_{\perp}=8.0(5)\times10^{-4}$ emu/mol (Fig.\:2a), the specific heat coefficient of the fermionic excitations, which corresponds to the Sommerfeld constant in metals, is estimated to be $\approx 56$\,mJ/K$^2$mol. This value is of the same order of magnitude reported by the heat capacity measurements~\cite{Yamashita11}. Moreover, the Fermi temperature, which is predicted to be the order of magnitude of $J/k_B$($\sim250$\,K)~ \cite{Katsura10}, is estimated to be $\sim 480$\,K. Thus, although the present simple estimations should be scrutinized, these thermodynamic quantities appear to be quantitatively consistent with the theory of the QSL that possesses a spinon Fermi surface.   In d$_9$-dmit, while $\gamma$-value is enhanced by nearly twice~\cite{Yamashita11}, $\chi_{\perp}$ is reduced by nearly 15\%, compared with h$_9$-dmit.   This may indicate that the nonmagnetic excitations, which enter not in $\chi_{\perp}$ but in $\gamma$, are enhanced  close to the magnetic end point, whose origin deserves further studies.

The most direct evidence of the presence of the Fermi surface may be given by the quantum oscillation measurements \cite{Motrunich06}.   To search such oscillations, we have measured the torque up to $\mu_0H=32$\,T at 30\,mK (Supplementary Information). We find an extended paramagnetic response up to the highest field, but no discernible oscillation is observed within the experimental resolution (inset in Fig.\:2b).  A possible explanation for this is that the coupling between the applied magnetic field and the gauge flux which causes the quantum oscillations is very small. 

The presently revealed extended Pauli-paramagnetic quantum phase bears striking resemblance to metals with Fermi surface rather than insulators even though the charge degrees of freedom are frozen. We also emphasise that the presence of such a novel quantum phase might  provide profound implications on the physics of other class of strongly interacting many-body systems in the vicinity of the end point of long-range order as well as the exotic quantum spin systems.

\begin{acknowledgements}
 We thank fruitful discussion with L. Balents, S. Fujimoto, T. Goto, M. Imada, N. Kawakami, Y.\,B. Kim,  P.\,A. Lee, R. Moessner, N. Nagaosa, T. Sasaki, T. Senthil, and K. Totsuka.  This research has been supported through Grant-in-Aid for the Global COE program ``The Next Generation of Physics, Spun from Universality and Emergence'' from MEXT of Japan, KAKENHI from JSPS,  and grant-in-aid for Scientific Research on Innovative Areas (No. 20110003)
from the Ministry of Education, Culture, Sports, Science and Technology (MEXT).
\end{acknowledgements}

\section{Supplementary Information}

\subsection{Methods}

Single crystals of {\sldmit}
(Et = C$_2$H$_5$, Me = CH$_3$, dmit = 1,3-dithiole-2-thione-4,5-dithiolate) were grown by the air-oxidation method. This material has a layered structure (space group $C2/c$) with the lattice parameters $a=14.515$\,\AA, $b=6.408$\,\AA, and the inter-layer distance $d=c^*/2=18.495$\,\AA. The deuterated crystals of d$_9$-EtMe$_3$Sb[Pd(dmit)$_2$]$_2$ were prepared by the same method using (Et(CD$_3$)$_3$Sb)$_2$[Pd(dmit)$_2$]$_2$. The deuterium atoms were introduced by CD3I (99.5\%D). Typical sample size is $\sim 1$\,mm $\times$ 1\,mm $\times$ 0.05\,mm. 

All magnetic torque measurements were carried out by attaching one single crystal to a cantilever with a tiny amount of grease. Piezo-resistive cantilevers were used in the measurements at Kyoto University ($\mu_0H < 7$\,T, $T > 0.3$\,K) and at NIMS ($\mu_0H < 17$\,T, 30\,mK $\leq T \leq 1$\,K). In the measurement at High magnetic field facility at Grenoble (30\,mK $\leq T \leq 1$\,K), magnetic fields up to 32 T were provided by a resistive magnet and a capacitance-sensing metallic cantilever was used. To determine the absolute value of $\chi_\perp$  (Fig.\:2a), sensitivities of the each cantilever were calibrated in situ by detecting $\sin\theta$ oscillation due to the sample mass at zero field. Thermal equilibrium between samples and cryostat at the lowest temperature was ensured by immersing the whole setup into the mixture of the dilution refrigerator.
 
Electron paramagnetic resonance (EPR) measurements have been performed in RIKEN by using a conventional X-band EPR spectrometer (JEOL, JES-RE3X) equipped with a continuous He-flow cryostat. The angular dependence measurement of the $g$-value was carried out using the same single crystal at 4\,K, and the $g$-value for each angle has been precisely obtained from the correction using the Mn$^{2+}$ field marker.

\subsection{Determination of susceptibility from the torque measurements}\label{chi}

Here, we describe how we determine the uniform susceptibility $\chi_{\perp}$ and magnetization $M_{\perp}$ from the torque measurements.

By adopting principal axes of $g$-factor ($\bm{p}$, $\bm{q}$ and $\bm{r} = \bm{p} \times \bm{q}$ as shown in Fig.\:1c and d) as orthogonal coordinate system, the spin susceptibility tensor can be diagonalized as
\begin{equation}
\bm{M} = \begin{pmatrix}
	\chi_{pp} & 0 & 0 \\
	0 & \chi_{qq} & 0 \\
	0 & 0 & \chi_{rr}
\end{pmatrix}
 \bm{H}. \label{chidiag}
\end{equation}

The torque detected by the cantilever method is a component perpendicular to the field-rotating plane (Fig.\:1b). For the $c^\ast-a$ rotation this is given as 
\begin{eqnarray}
\tau_{c^\ast-a}(\theta) &=& \left( \bm{c^\ast} \times \bm{a} \right) \cdot \left( \bm{M} \times \bm{H} \right)	\\
&=& \frac{1}{2}\mu_0H^2V(\chi_{qq}-\chi_{pp})\sin 2(\theta+\theta_0),
\end{eqnarray}
and for the $c^\ast-b$ rotation we obtain 
\begin{eqnarray}
\tau_{c^\ast-b}(\theta) &=& \left( \bm{c^\ast} \times \bm{b} \right) \cdot \left( \bm{M} \times \bm{H} \right) \\
&=&\frac{1}{2}\mu_0H^2V(\chi_{pp}\sin^2\theta_0+\chi_{qq}\cos^2\theta_0-\chi_{rr})\sin 2\theta. \nonumber\\
\end{eqnarray}

These equations show that the torque should have a sinusoidal shape with two-fold oscillation crossing zero at the maximum and minimum positions of the $g$-factor. This is exactly what we observed (Fig.\:1:e and f). This indicates that the susceptibility can be diagonalized as in Eq.\:(\ref{chidiag}) and that the anisotropy of the $g$-factor is the origin of the anisotropy of the susceptibility in this spin liquid system. 

Having established that the magnetic susceptibility reflects the anisotropy of $g$-factor, we assume that the susceptibility is proportional to the square of the $g$-factor ($\chi_{ii} = g_{ii}^2 \tilde{\chi}$) as in conventional paramagnetic systems. Then $\chi_{ij}$ can be obtained by the reduced susceptibility $\tilde{\chi}$, which is determined from $\tau_{c^\ast-a}$ and $\tau_{c^\ast-b}$ as
\begin{eqnarray}
\tau_{c^\ast-a}(\theta) &=& \frac{1}{2}\mu_0H^2V(g_{qq}^2-g_{pp}^2)\tilde{\chi}\sin 2(\theta+\theta_0) \\
\tau_{c^\ast-b}(\theta) &=& \frac{1}{2}\mu_0H^2V(g_{pp}^2\sin^2\theta_0+g_{qq}^2\cos^2\theta_0-g_{rr}^2)\tilde{\chi}\sin 2\theta \nonumber\\
\end{eqnarray}
where the principal values of $g$-factor and $\theta_0$ are provided by the electron paramagnetic resonance measurements (Fig.\:1e and f) as $g_{pp} = 1.9963$, $g_{qq} = 2.0325$, $g_{rr} = 2.0775$, $g_{c^\ast c^\ast} = 2.0217$ and $\theta_0 = 32$\,degree.
The susceptibility for the magnetic field perpendicular to the basal plane is, therefore, given by
\begin{eqnarray}
\chi_{\perp} = g_{c^\ast c^\ast}^2 \tilde{\chi}, \qquad M_{\perp} = \chi_{\perp}H.
\end{eqnarray}
The two independent measurements $\tau_{c^\ast-a}$ and $\tau_{c^\ast-b}$ give quantitatively consistent results for $\chi_{\perp}$ (red and green circles in Fig.\:2a, respectively), confirming the validity of our analysis. 

\subsection{Origin of the paramagnetic susceptibility}

The obtained $\chi_{\perp}(T)$ in the spin-liquid state of EtMe$_3$Sb[Pd(dmit)$_2$]$_2$ is nearly temperature-independent at low temperatures below $\sim 2$\,K. This immediately indicates that in the torque measurements, the Curie contribution from the free spins of the impurities is almost exactly cancelled out. This indicates that such free spins are not subject to the $g$-factor anisotropy, which may be explained if the impurity spins reside at the EtMe$_3$Sb cation site between the dmit layers. The absence of the Curie term is also confirmed from the field-linear magnetization without saturation in a wide field range covering large $B/T$ ratios, which cannot be fitted by the Brillouin function. 

The paramagnetic temperature-independent susceptibility, in general, may be originated from the orbital Van Vleck susceptibility and the spin Pauli susceptibility. However, the Van Vleck term is found to be negligibly small in the closely related isostructural material Et$_2$Me$_2$Sb[Pd(dmit)$_2$]$_2$ with slightly different cations Et$_2$Me$_2$Sb. In this material, a large spin gap is formed below the first-order transition at $\sim 70$\,K, which has been attributed to the charge disproportionation in [Pd(dmit)$_2$]$_2$ anions. The susceptibility jumps from the value close to the present EtMe$_3$Sb[Pd(dmit)$_2$]$_2$ case at high temperature to almost zero below the transition, indicating the absence of the Van Vleck paramagnetic susceptibility \cite{Tamura05}. This strongly suggests that the sizable paramagnetic temperature-independent susceptibility in the spin liquid state of EtMe$_3$Sb[Pd(dmit)$_2$]$_2$ can be fully attributed to the spin Pauli susceptibility. We also note that the diamagnetic Landau susceptibility, if present, should give a very minor correction to the Pauli susceptibility, because we did not observe the quantum oscillations at high magnetic fields. 

\subsection{Estimation of thermodynamic quantities of 2D Fermion from $\chi_{\perp}$}
Here we explain the estimations of the specific heat coefficient ($\gamma$) and the Fermi temperature ($T_F$) by assuming the system can be described by simple 2D fermions with Fermi surface as in conventional metals, even though the system is an insulator.  The Pauli susceptibility is given by 
\begin{equation}
\chi_{\perp}=\frac{1}{4}g_{c^{\ast}}^2 \mu_B^2 D(\varepsilon_F),
\end{equation}
where $\varepsilon_F$ is the Fermi energy, $D(\varepsilon_F)$ is the density of states and  $\mu_B$ is the Bohr magneton.  In 2D systems, $D(\varepsilon_F)=n/\varepsilon_F$ where $n$ is the number of fermion (spin) per volume.  Using $\chi_{\perp}=8.0(5)\times10^{-4}$ emu/mol (Fig.\:2a), the specific heat coefficient of the fermionic excitations, which corresponds to the Sommerfeld constant in metals, yielded by the equation 
\begin{equation}
\gamma = \frac{1}{3} \pi^2 k_B^2 D(\varepsilon_F) = \frac{1}{3}\pi^2k_B^2 \frac{4 \chi_{\perp}}{g_{c^{\ast}}^2 \mu_B^2} \sim \text{56~mJ/K$^2$ mol}
\end{equation}
This value is of the same order of magnitude as the values reported by the heat capacity measurements~\cite{Yamashita11}. 
Provided that  a substantial portion of the gapless excitations observed by specific heat is magnetic, the $\gamma$ and $\chi_{\perp}$ values result in a Wilson ratio close to unity, which is a basic properties of metals. 
The Fermi temperature, given by  the relation  
\begin{equation}
T_F=\varepsilon_F/k_B=\frac{g_{c^{\ast}}^2\mu_B^2}{4\chi_{\perp}k_B} \sim \text{480\,K},
\end{equation}
which is in the same order as $J/k_B \sim 250$~K.

\subsection{Search for quantum oscillations}

To find a quantum oscillation due to a spinon Fermi surface, the magnetic torque of the pristine sample (h$_9$-dmit (C)) was measured up to 32\,T. The magnetization $M_{\perp}$ is estimated from the torque as described in section \ref{chi}. After subtracting the background signal determined by a linear approximation to data above 12\,T, the data is plotted as a function of $1/H$ as red line in the upper inset of Fig.\:2b. However, no discernible oscillation is observed within the experimental resolution.

An important requirement to observe the quantum oscillation in the torque (de Haas-van Alphen effect) in a metal is a long mean free path $\ell=v_F\tau$ ($v_F$ is the Fermi velocity and $\tau$ is the scattering time) to satisfy that $\omega_c\tau=\frac{eB\tau}{m^*}=\mu B$ exceeds unity, where $\omega_c$ is the cyclotron frequency, $m^*$ is the effective mass and $\mu$ is the mobility. In a simple metal, the mobility can be estimated by using the residual thermal conductivity $\kappa_0/T$, the Sommerfeld constant $\gamma$, and the Fermi energy $\varepsilon_F$ through
\begin{equation}
\mu=\frac{e\tau}{m^*}=\frac{3\kappa_0/T}{2\gamma\varepsilon_F}.
\end{equation}
By using $\kappa_0/T\approx0.2$\,W/K$^2$m \cite{Yamashita10}, $\gamma\approx 20$\,mJ/K$^2$mol \cite{Yamashita11}, and $\varepsilon_F/k_B\sim 480$\,K estimated by the present susceptibility measurements, we can roughly estimate $\mu\sim 0.18$\,T$^{-1}$ in the spin-liquid state of EtMe$_3$Sb[Pd(dmit)$_2$]$_2$. This yields $\omega_c\tau>5$ at 32\,T, which strongly suggests that the impurity scattering effect is not the main origin of the absence of the quantum oscillations. 

A possible explanation for the absence of the quantum oscillations is that the coupling between the applied magnetic field and the gauge flux which causes the quantum oscillations is very small. This seems to be consistent with the fact that our torque curves can be described only by the $g$-factor anisotropy, without showing any contribution from the orbital (spinon) motion in the 2D plane. We also point out that the thermal Hall effect in this material \cite{Yamashita10} is negligibly small compared with a theoretical suggestion \cite{Katsura10}, which may be originated from the same physics.


\end{document}